\documentclass[twocolumn,amssymb,floatfix,aps,prl,showpacs,graphicx]{revtex4}
\usepackage{epsfig} 
\usepackage{amsmath}
\usepackage{times}

\begin{document} 

\title{ Exponential
Orthogonality Catastrophe at the Anderson Metal-Insulator Transition}

\author{S. Kettemann} 

\address{Jacobs University, School of Engineering and Science,
  Campus Ring 1, 28759 Bremen, Germany}
\address{ Division of Advanced
  Materials Science Pohang University of Science and Technology
  (POSTECH) San 31, Hyoja-dong, Nam-gu, Pohang 790-784, South Korea}

%
%
%
%

\date{\today}

\begin{abstract} 
 We consider the   orthogonality catastrophe
 at the  
  Anderson Metal-Insulator transition (AMIT). 
 The typical overlap $F$
    between the ground state 
      of a Fermi liquid and the one 
       of the same system with an added potential impurity 
is found to  decay at the AMIT
 exponentially with system size $L$  as $F \sim \exp (- \langle I_A\rangle /2)=  \exp(-c L^{\eta})$, where
  $I_A$ is the so called Anderson integral,  $\eta $
   is the power of multifractal intensity correlations and $\langle ... \rangle$ denotes 
    the ensemble average. Thus,  strong 
     disorder  typically   increases the sensitivity of a system to an additional impurity exponentially.
    We recover on  the metallic side of the transition 
     Anderson's result that  fidelity $F$ 
       decays with a power law $F \sim L^{-q (E_F)}$ with system 
        size $L$.
     This power increases as  Fermi
         energy $E_F$ 
          approaches mobility edge $E_M$ as $q (E_F) \sim  (\frac{E_F-E_M}{E_M})^{-\nu \eta},$
  where $\nu$ is the critical exponent of  correlation length $\xi_c$. 
     On the insulating side of the transition $F$
      is constant for system sizes exceeding  localization 
       length $\xi$. While these results  are obtained from the  
        mean value of $I_A,$ giving  
       the  typical fidelity $F$, we  find that 
         $I_A$ is widely, log normally, distributed 
          with a width  diverging at the AMIT. As a consequence, the mean value 
           of fidelity $F$  converges to one at the AMIT, in strong contrast to 
            its typical value which converges to zero exponentially fast with 
             system size $L$. This counterintuitive behavior is explained 
             as  a  manifestation of 
              multifractality at the AMIT. 
\end{abstract}



\maketitle


 Anderson showed  in Ref. \onlinecite{ao}
 that the addition of  a static potential impurity to 
  a system of N fermions changes  its groundstate such that 
   the overlap 
   between  the original  $\langle  \psi \mid$ and the new ground state
   $\langle  \psi' \mid$  has an upper bound,
   \begin{equation}
    \chi = F^2 =
     |\langle  \psi \mid \psi' \rangle|^2 < \exp \left(- I_A\right),
   \end{equation}
   where the Anderson integral $I_A$ is 
    for noninteracting electrons given in terms 
     of the single particle eigenstates of the original system 
     $|n\rangle$ and the new system $|m'\rangle$ by 
      \begin{equation}
I_A = \frac{1}{2} \sum_{\epsilon_n \le 0, \epsilon_{m'}>0}
|\langle n | m' \rangle|^2.
            \end{equation} 
 If the added   impurity is short ranged and 
  of strength $\lambda$, Anderson found 
   for a clean metal $I_A = 2 \pi^2 \lambda^2 \ln N,$
 diverging with the number of fermions $N$, so that $F$, also called
  fidelity, decays with a power law with  $N,$  leading to  the so called
  orthogonality catastrophe (AOC). This implies that the local perturbation 
  connects the system to a macroscpic number  of excited states which has 
   important consequences like the  singularities in the X-Ray absorption and emission of metals\cite{nozieresdominicis}.   Furthermore,  the zero bias anomaly  in
   disordered metals\cite{altshuleraronov} and 
   anomalies in the tunneling density of states in quantum Hall systems\cite{tunnelingDOS}
   are related to the AOC
   The concept of fidelity can be generalised
    to any parametric perturbation of a system and be used to 
     characterise quantum phase transitions \cite{venuti}.
      The AOC has been explored
  in  mesoscopic systems\cite{meso1,meso2}. 
     With the advent  of 
     engineered many-body systems in ensembles of ultracold atoms it is possible to study nonequilibrium quantum dynamics of such systems in a 
      controlled way so that conseuqences of 
      parameter changes become measurable directly\cite{demler}.
     
An intriguing question is, if the system becomes less or more  sensitive to the 
    addition of another impurity if it already contains a finite density of impurities. 
Gefen et al. showed in Ref. \onlinecite{gefenlerner} that 
    in a weakly disordered metal the average 
     value $\langle I_A\rangle $   scales with $\ln N$
       when the potential of the 
        added impurity potential is short ranged. 
 That  result is valid to leading order in 
          $1/g,$   where $g = k_{\rm F} l_e \gg 1$ is a measure of disorder strength
    with   mean free path $l_e$.
           Numerical results \cite{gefenlerner}
           show  that in a 2-dimensional 
            disordered system $I_A$ increases as  the 
            disorder strength  increases 
            until it  is so strong 
              that  localization length $\xi$ is smaller than 
               system size $L$.   Beyond that,    $I_A$ decreases 
               as  $\xi$ decreases with increasing disorder. 
               Thus, the addition of an impurity changes the ground state 
                of weakly disordered systems more strongly than the one of  a clean system. 
             Only at strong disorder when the fermions are localized, 
             its sensitivity to a potential change  decreases again.

\begin{figure}[t]
\includegraphics[width=7cm,angle=0]{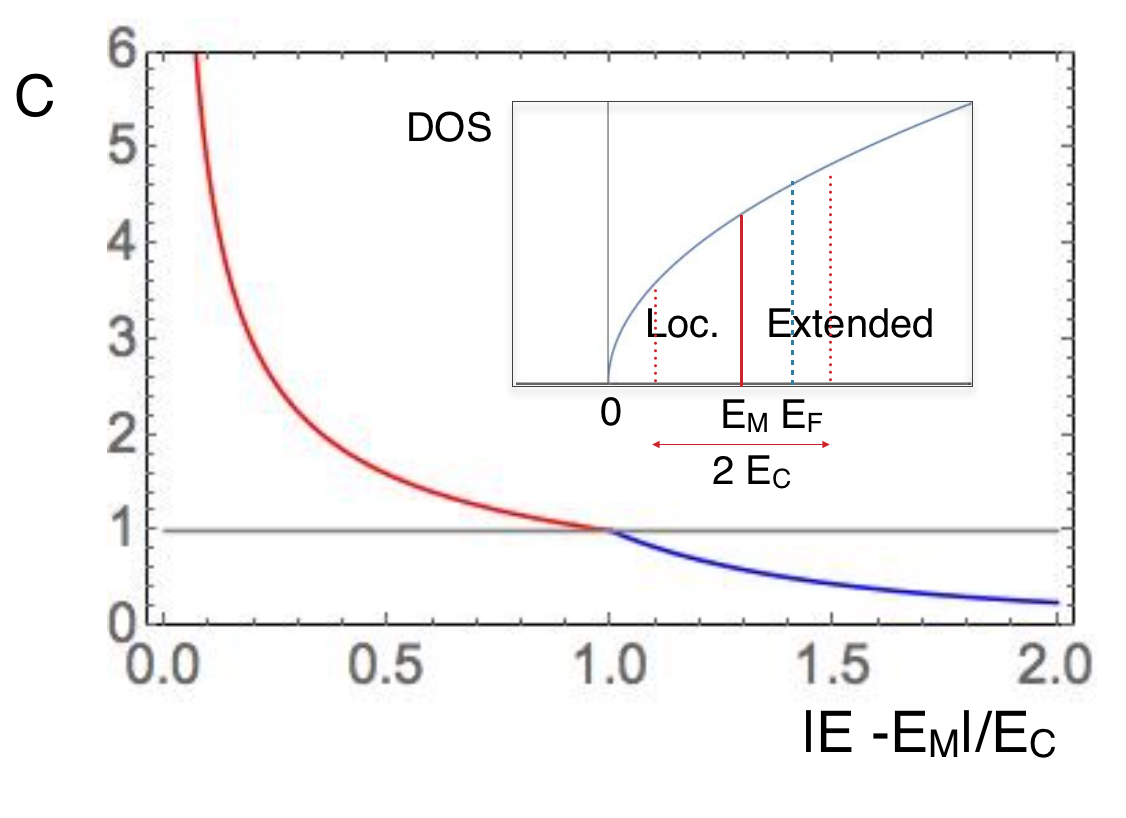}
\caption{Correlation function of intensities $C$ 
as function of their energy difference $|E-E_M|$ in 
  a  disordered system with $\eta =2,d=2$. Inset: Density of states (DOS) 
   as function of energy $E$  with a transition 
    between localized and extended states at mobility edge $E_M$. 
     Intensities are correlated within an energy window  $2E_c$  around $E_M$.}
\label{fig:c}
\end{figure}

        Here, we aim to derive  this behavior analytically in order to find out 
        how fidelity $F$ changes when tuning disorder
         strength or  energy. 
         In systems  close to the Anderson metal-insulator transition (AMIT)
         we  can use
           the fact that the single particle wavefunctions 
            at the AMIT  are multifractal \cite{multifractal} and power law correlated\cite{powerlaw,cuevas,ioffe}. 
              We also obtain  analytical results for noncritical 2-dimensional 
               disordered systems, which confirm the numerical calculations of 
                Ref. \onlinecite{gefenlerner}.  
        For a short range impurity of strength $\lambda$, located at 
        position ${\bf x},$  $I_A$ can be expressed 
          in terms of the local  intensities of the unperturbed 
           Eigenstates $|n>$ with Eigenenergies
           $E_n,$  \cite{ao,gefenlerner} 
         \begin{equation} \label{ia}
         I_A = \frac{(2 \pi \lambda)^2}{2 \rho^2}
         \sum_{E_n < E_F}  \sum_{E_{m} > E_F}
          \frac{| \psi_n({\bf x})|^2 | \psi_m({\bf x})|^2  }{(E_n-E_m)^2},
         \end{equation}
         where 
          $\rho$ is the mean density of states. 
          The correlation function of the
intensities associated to two energy levels distant by
$\omega_{nm} =E_n- E_m$ is   given by
\cite{cuevas,ioffe,ioffe2} 
\begin{eqnarray}
\label{cc}
 && C(
\omega_{nm} = E_n-E_m) = L^d \int d^dr\, \left\langle |\psi_n({\bf r})|^2
 |\psi_m({\bf r})|^2 \right\rangle \nonumber \\ &&= \left\{
 \begin{array}{ll}(\frac{E_c}{{\rm Max} (|\omega_{nm}|, \Delta) })^{\eta/d}, 
 & 0 < |\omega_{nm}| < E_c, \\ (E_c/|\omega_{nm}|)^{2}, &
|\omega_{nm}| > E_c,
\end{array}
\right.,
\end{eqnarray}
when $E_n \le E_M$ and $E_m \ge E_M$ or vice versa\cite{kats}.    $\Delta = 1/(\rho L^d)$ is 
the average level spacing.
Here,  $\eta = 2 (\alpha_{0}-d)$, with multifractality parameter $\alpha_0$  
 and $d$ the dimension. 
 The correlation energy $E_c$ is a macroscopic energy of order
  of  elastic scattering rate $1/\tau$.
For
$|\omega_{nm}| < E_c$ correlations are enhanced in comparison to the
plane-wave limit $C_{nm}=1$, see Fig. \ref{fig:c}, 
where we set one of the energies at he mobility edge $E_n=E_M$
 and the other at $E_m=E$. Note, that for $|\omega_{nm}| > E_c$
 it decays below $1$. This anticorrelation 
  ensures that the total  intensity at a position ${\bf x}$ is normalised: A dip in  intensity 
   at one energy implies an enhancement of intensity at another energy 
    and vice versa. 
                                     
          {\it Mean Value of  the Anderson Integral.---}
          Inserting Eq. (\ref{cc}) into Eq. (\ref{ia}),
         the  average  mean value of  $I_A$ 
            is
                  \begin{equation} \label{iafm}
       \langle  I_A \rangle = \frac{(2 \pi \lambda)^2}{2 }
       \hspace{-.5cm}\iint\limits_{\epsilon < - \Delta/2,
        \epsilon' >  \Delta/2}\hspace{-.5cm}d \epsilon d \epsilon'
          \frac{C_{\epsilon,\epsilon'}  }{(\epsilon-\epsilon')^2}.
         \end{equation}
          This  gives  the geometrical average of $F,$
         $\exp (\langle \ln F \rangle)=\exp (-   \langle  I_A \rangle /2 )$.
         At the AMIT we  get 
           with Eq. (\ref{cc}) 
           \begin{equation} \label{iafmEM}
       \langle  I_A \rangle|_{ E_F = E_M} = \frac{2 ( \pi \lambda)^2}{\gamma (1+\gamma) }
      \left(  \frac{E_c}{ \Delta} \right)^{\gamma},
         \end{equation}
        diverging with  number of  particles $N = E_F/\Delta$
          with a power law,  $\gamma = \eta/d$.

           As the Fermi energy is moved into the insulating regime $ E_F < E_M$, 
            there remain  multifractal correlations, Eq. (\ref{cc}),
             but the integral is now cut off at  local level spacing 
              $\Delta_{\xi} = 1/(\rho \xi^d)$, since there is local level repulsion\cite{kats}. This yields
               \begin{equation} \label{iafmI}
       \langle  I_A \rangle|_{E_F <E_M} = \frac{2 ( \pi \lambda)^2}{\gamma (1+\gamma) }
      \left(  \frac{E_c}{ \Delta_{\xi} } \right)^{\gamma},
         \end{equation}
     independent of the number of particles  $N$.

      In the metallic regime $ E_F > E_M$ 
            all wave functions are extended.
 On length scales
smaller than correlation length $\xi_c$ multifractal fluctuations
  still occur and
there are  power-law correlations
in energy Eq. 
 (\ref{cc}).
            The energy difference
$|\omega_{nm}|$ is for $\epsilon_{n} < \epsilon_{m}$  substituted by ${\rm Max} [|\omega_{nm}|,  
\Delta_{\xi_{c n}}]$, where $\Delta_{\xi_{c n}} = E_{c}\,
(\xi_{c n}/a_{c} )^{-d}$. $\xi_{c n}$ is the correlation  
 length at energy $E_n$ and $a_{c}$ is a small length scale defined by $E_{c}=1/(\rho
a_{c}^d)$\cite{kats}. 
 For
$\omega_{nm} < E_c$, correlations are enhanced in comparison to
plane-wave limit $C_{nm}=1$,
yielding
               \begin{eqnarray} \label{iafmmetal}
   \langle  I_A \rangle|_{E_F >E_M} = 2 ( \pi \lambda)^2 \left(  \frac{E_c}{ \Delta_{\xi_c}} \right)^{\gamma} 
       \left( \frac{1}{\gamma (1+\gamma) }   + \ln \frac{N}{N_{\xi_c}} \right),
         \end{eqnarray}
         diverging logarithmically with  number of electrons $N$ in agreement with 
          Anderson's result for a metal, which is recovered exactly far away from the MIT, 
           where 
          $\Delta_{\xi_c} \rightarrow E_c$.

\begin{figure}[t]
\includegraphics[width=8.5cm,angle=0]{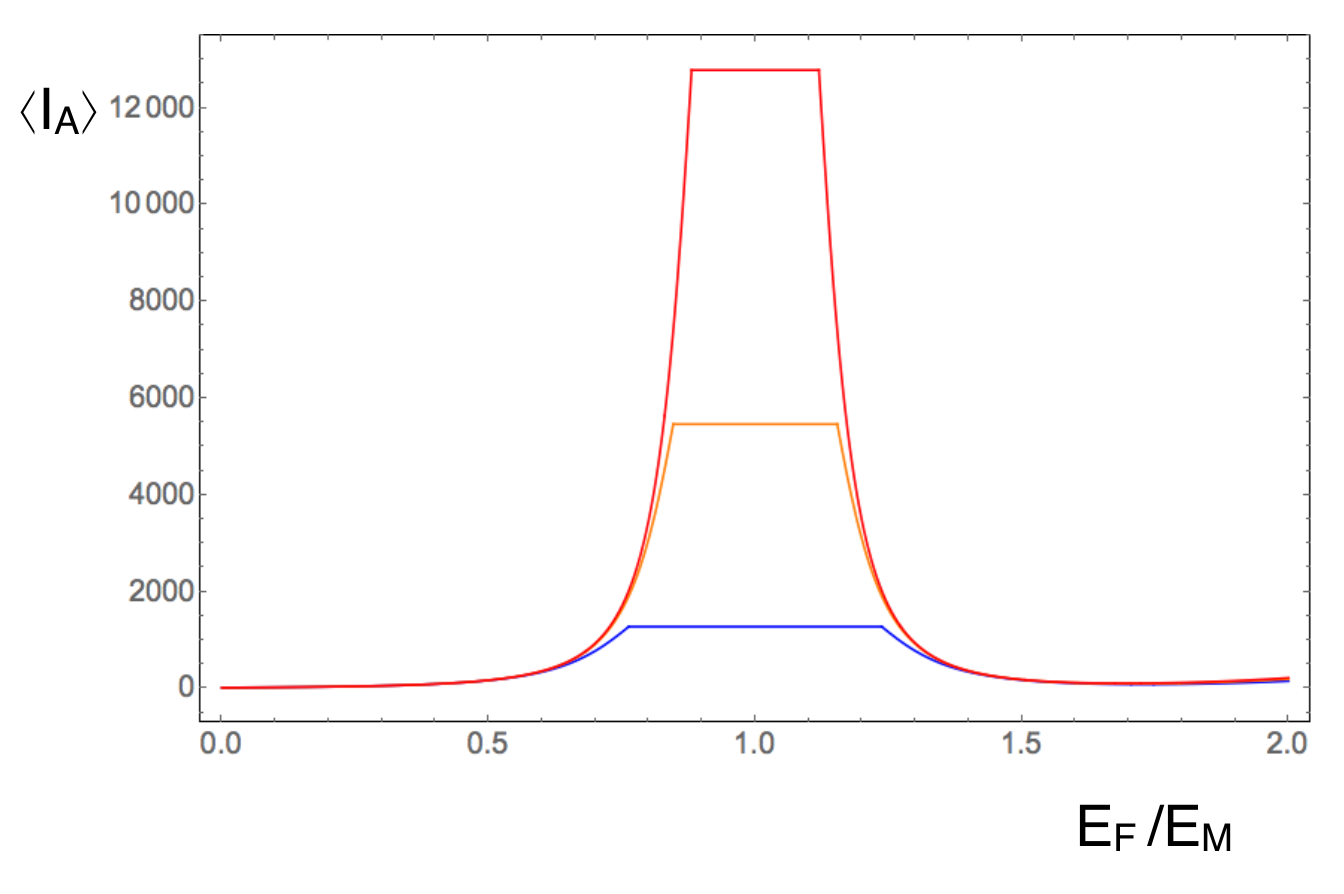}
\caption{The  average Anderson integral  $\langle I_A \rangle$ as function of Fermi energy $E_F$ in 
units of  mobility edge $E_M$ for a 
 3-dim. disordered system,
  Eqs. (\ref{iafmEM},\ref{iafmI},\ref{iafmmetal}), for various system 
  sizes $L$ (in units of microsopic length $a_c$),  $L=10$ (blue), $L=20$ (orange) and $L=50$ (red).}
\label{fig:ai}
\end{figure}

  In Fig. \ref{fig:ai} we plot the first moment of the  Anderson Integral  $  \langle  I_A \rangle$ as function 
   of Fermi energy $E_F$ in units of mobility Edge $E_M$,
    Eqs. (\ref{iafmEM},\ref{iafmI},\ref{iafmmetal}), for various system
     sizes L.

\begin{figure}[t]
\includegraphics[width=8.5cm,angle=0]{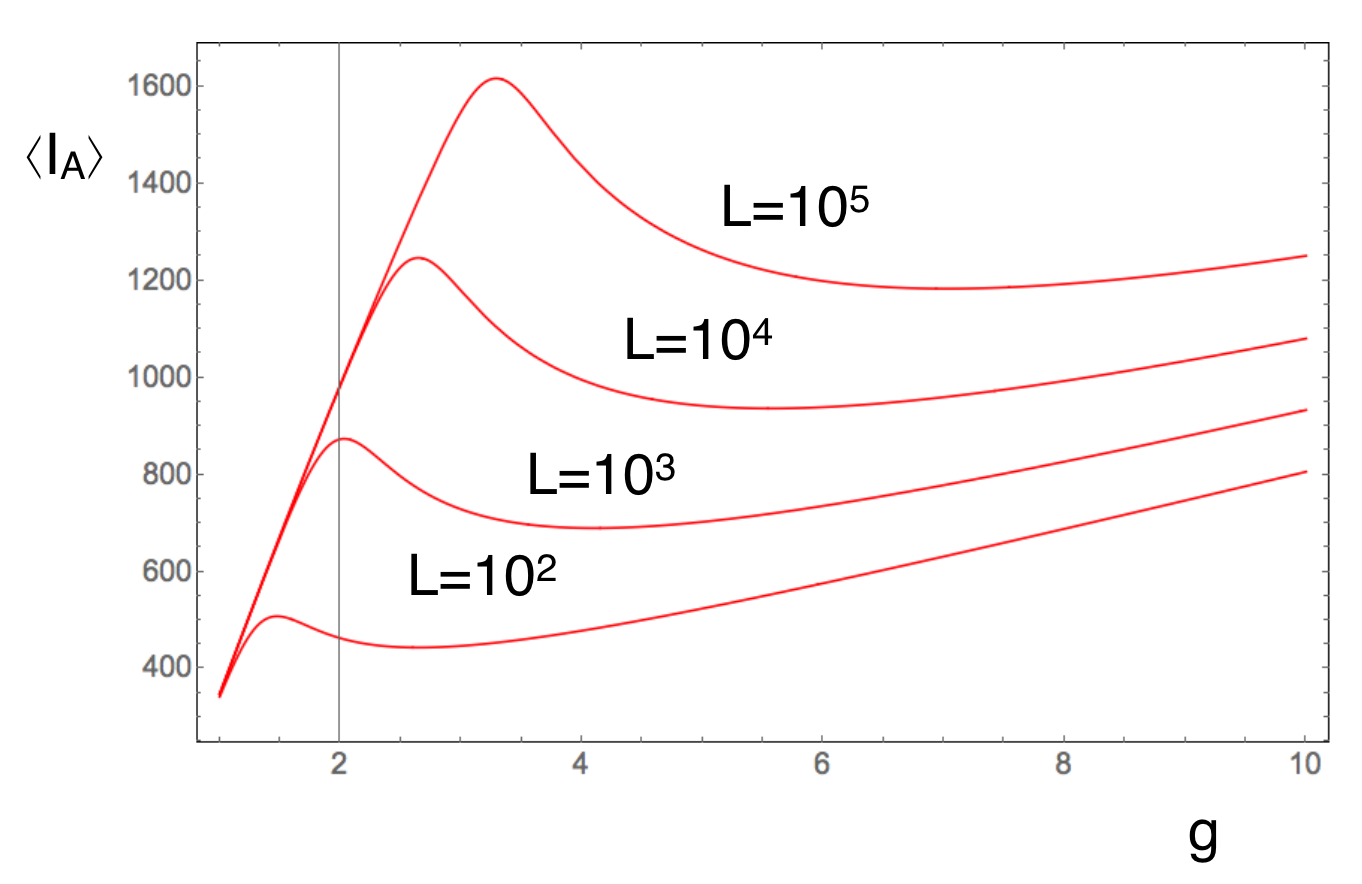}
\caption{The average Anderson integral  $\langle I_A \rangle$ as function of 
disorder paramter $g= E_F \tau$  for a
 2-dimensional disordered system, as obtained by  
  substituting 
                  $\xi_{2D} = g \exp (\pi g)$ and  $ \gamma_{2D} = 1/(\pi g)$ into Eq. 
                    (\ref{iafmI}), 
                    with $\Delta_{\xi} = D/\xi^2$ and $\Delta = D/L^2$
 for various system 
  sizes $L$.}
\label{fig:2d}
\end{figure}               
          {\it Anderson Integral of 2D Disordered Electon Systems.---}
           In 2D disordered electron 
           systems without spin-orbit interaction, 
            without strong magnetic field and strong interactions all state
           are localized. Therefore, the Anderson integral 
     is given by Eq. (\ref{iafmI})
            with localization length $\xi_{2D} = g \exp (\pi g)$ (in units 
             of the smallest microscopic length scale),  where
             $g = \epsilon_F \tau$. In 2D weakly disordered
              systems, $ g \gg 1,$ there are weak logarithmic 
              correlations of the intensity at different energies
              which can to leading order
                in $1/g$ be written as power law correlations 
                 with  power $ \eta_{2D} = 2/(\pi g)$. Substituting 
                  $\xi_{2D}$ and  $ \gamma_{2D} = 1/(\pi g)$ into Eq. 
                    (\ref{iafmI}), 
                    with $\Delta_{\xi} = D/\xi^2$ and $\Delta = D/L^2$, 
                     where $D$ is the band width, 
                    we get the Anderson integral for 2D disordered systems
                     as function of disorder paramter $g$ and system size $L$ as plotted in Fig. \ref{fig:2d}.  We thus confirm analytically the 
                     nonomonotonous dependence
                      of   $\langle  I_A \rangle$ as function of disorder
                       strength as observed numerically in 
                       2D disordered systems in  Ref. \onlinecite{gefenlerner}.
                        We find that the  maximal value of $I_A$ increases
                         with system size logarithmically. 
                       When the localization length exceeds the system size $\xi >L$
                      we find   $  \langle  I_A \rangle = 2 \pi (\pi \lambda)^2
                      g  L^{2/(\pi g)}$. Expanding in $1/g$
                       we recover to leading order Anderson's result, 
                       $  \langle  I_A \rangle =const. + 2 ( \pi \lambda)^2
                      \ln L^2$. Note that  the logarithmic divergence   is independent of the disorder paramter $g$ and coincides exactly with Anderson's result for a clean metal. 
                  In the opposite limit, when localization length $\xi$
                   is smaller than systems size $L$, we find    $  \langle  I_A \rangle = 2 \pi (\pi \lambda)^2
                      g  \exp (2 + 2 \ln g/(\pi g)) $.

           {\it Distribution Function  of  the Anderson Integral.--- }
             Having obtained that the average Anderson integral 
              diverges with the system size $L$ at the AMIT more strongly 
               than in a clean or weakly disordered metal, 
                we may ask how widely   $I_A$ is distributed. 
                 Since it is a functional of the local density of states
                  at the position where 
                  the additional impurity has been placed,
                  Eq. (\ref{ia}), it
                   is expected to be widely distributed. 
             The distribution function of 
              $I_A({\bf x})$ as obtained by placing the impurity 
               in different ensembles with  same disorder strength
               is  
               \begin{equation} \label{d}
                P(I_A) = \int \prod_l d \alpha_l  P\left(\{  \alpha_l \} \right)
                \delta \left( I_A - I [\{  \alpha_l \}  ] \right),
               \end{equation}
             where  $I [\{  \alpha_l \}  ]$
              is defined by the right side of Eq. (\ref{ia}).
               Following the strategy recently used  in 
  the  derivation of the distribution function  of
 Kondo temperatures  $T_{K}$  at the AMIT\cite{kats}, we
    replace the correlated distribution function of 
     all intensities $P\left(\{  \alpha_l \} \right)$
      by a product of pairwise 
joint distribution functions of $ |\psi_n({\bf r}_{i})|^2$ and
$|\psi_m({\bf r}_{i})|^2$, in accordance with  the correlation
function   $C_{nm}$, Eq. (\ref{cc}).
   Thereby,  we can derive the conditional
intensity of a state at energy $E_{l}$, given that  the intensity at
mobility edge  $E_{M}$ is $|\psi_M ({\bf r})|^2 =
L^{-\alpha}$,\cite{kats}
\begin{eqnarray}
\label{cci}
L^d \langle| \psi_{l} ({\bf r}) |^2
\rangle_{|\psi_M({\bf r})|^2=L^{-\alpha}} =  \left|
\frac{E_{l} -E_{M} }{E_{c}} \right|^{r_{\alpha}},
\end{eqnarray}
where the power is given by
$r_{\alpha}=\frac{\alpha-\alpha_{0}}{d} -
  \frac{  \eta }{2 d } g_{lM}$.
 When $E_{l}$ is located  away from
mobility edge $E_{M}$, the coefficient $g_{lM} = \ln | (E_{l}
-E_{M}) /E_{c} |/ (d \ln L) $ vanishes for $L \rightarrow \infty$. 
Close to $E_{M}$ it saturates: $ g_{lM}
|_{E_{l}\rightarrow E_{M}} \rightarrow -1 $ and Eq. (\ref{cci})
reduces to $L^{d-\alpha}$, the local intensity at $E_{M}$ relative to
the intensity of an extended state $L^{-d}$\cite{kats}.
At positions where the local  intensity at the mobility edge
is small, corresponding to  $\alpha >\alpha_{0}$, it
is suppressed within an energy range of order $E_{c}$ around $E_{M},$ forming
 {\it   local pseudogaps } with  power $r_{\alpha} =
\frac{\alpha-\alpha_{0}}{d}$.
 When the
intensity at $E_{M}$ is larger than its typical value,
 $\alpha < \alpha_{0}$, there are {\it local power law divergencies} and the
local density of states  is enhanced within  energy range $2 E_{c}$ around
$E_{M}$, increasing as a power law when $E_{l}$ approaches $E_{M}$.
  

 Next, we  can 
  find the Anderson integral at a position ${\bf x}$, when
the intensity  of the state at the  mobility edge 
$E_{n}=E_{M}$ at that position
is fixed to $|\psi_n({\bf  x})|^2= L^{- \alpha},$ 
\begin{equation} 
     \label{iafm}
       \langle  I_A \rangle_{\alpha} = 2 (\pi \lambda)^2\hspace{-.5cm}\iint\limits_{\epsilon < - \Delta/2,
        \epsilon' >  \Delta/2}\hspace{-.5cm}d \epsilon d \epsilon' 
          \frac{1}{(\epsilon-\epsilon')^2} |\frac{\epsilon}{E_c}|^{r_{\alpha}} |\frac{\epsilon'}{E_c}|^{r_{\alpha}}.    
            \end{equation}
 which yields 
 \begin{equation} 
     \label{iafm}
       \langle  I_A \rangle_{\alpha} =   \frac{2 (\pi \lambda)^2}{\gamma (1+\gamma) }  \left(  \frac{E_c}{ D} \right)^{\gamma} 
       L^{-2(\alpha-d)}.
            \end{equation}
 Inverting this equation and inserting the 
  result into the distribution function of $\alpha$, $P(\alpha) = \exp [-\ln L
(\alpha-\alpha_0)^2 /(2   \eta)]$ we get with Eq. (\ref{d}),
  \begin{equation} 
     \label{p}
      P(  I_A  ) =   \frac{1}{\sqrt{8 \pi \eta \ln L}  }  \frac{1}{ I_A }
       e^{  - \frac{\left(  \ln \frac{I_A}{\langle I_A \rangle} + 2 \eta \ln L \right)^{2}}{8 \eta \ln L} },
            \end{equation}
            where $\langle I_A \rangle$ is the first moment,  Eq. (\ref{iafmEM}).
             Thus, the Anderson integral is  widely,  log-normally
              distributed with a width which increases with system size $L$ logarithmically. 
            
             If the Fermi energy is in the insulating regime, $E_F < E_M$, there
              is multifractality on length scales smaller than  localization length 
               $\xi$ and the intensity scales  with $\xi$, 
                  $| \psi_l({\bf x})|^2  = \xi^{-\alpha_l}.$ Thus, we find the distribution 
                   function of $I_A$ by replacing the system size $L$ by $\xi$ in the above 
                    derivation at the MIT yielding  Eq. (\ref{p}), where 
                     $L$ is replaced by $\xi$.
                    
               On the metallic side of the transition
              all wave functions are extended and their
intensities scale as $|\psi|^2 \sim L^{-d}$.
 On length scales
smaller than  correlation length $\xi_c$ multifractal fluctuations
of the wave function intensity  occur as long as $\xi_c$ is larger
than the microscopic length scale $a_{c}$.\cite{cuevas,ioffe2}
 In the metallic phase, moments of intensity
  scale  with  $\xi_c$ as 
  $L^{d q} \langle | \psi|^{2 q} \rangle \sim \xi_c^{(d-d_{q})(q-1)}$\cite{ioffe2,mirlin,kats}.
Therefore, we define  $\alpha$ in the metal as 
$L^d |\psi_{l} ({\bf r})|^2 =
\xi_{c l}^{d-\alpha_{l}}$,
 where $\xi_{c l}$ is the correlation length of
state $l$.  As  the MIT is approached $\xi_c$ diverges and is
replaced by system size $L$,  so that 
  $\alpha$ crosses over to the definition  used above at the MIT. 
 It has  to a good approximation
  the  Gaussian distribution, $
P(\alpha_{l}) \sim \exp [ - \ln
  \xi_{c l} (\alpha_{l}-\alpha_{0})^2/(2   \eta )]$ as confirmed numerically
   in Ref. \onlinecite{kats}. Therefore, in deriving the 
    distribution function of $I_A,$ we can follow the  strategy 
    used at the MIT, deriving first the value of $I_A$
     when averaged over all pair correlations, given that $\alpha$ at the
      Fermi energy is fixed, 
       \begin{equation} 
     \label{iafmm}
       \langle  I_A \rangle_{\alpha} =   2 (\pi \lambda)^2 c  \left( \frac{1}{\gamma (1+\gamma) }   + \ln \frac{N}{N_{\xi_c}} \right)
       \xi_c^{-2(\alpha-d)},
            \end{equation}
          where $c =  \left( E_c/ D \right)^{\gamma}$.
    Inverting this equation and inserting it into the distribution function of $\alpha$
  we get  with Eq. (\ref{d}) the distribution of $I_A$ in the metallic regime, Eq. (\ref{p}), 
                 replacing    $L$  by $\xi_c$
            and $\langle I_A \rangle$ by the
              first moment Eq. (\ref{iafmmetal}).

\begin{figure}[t]
\includegraphics[width=7.5cm,angle=0]{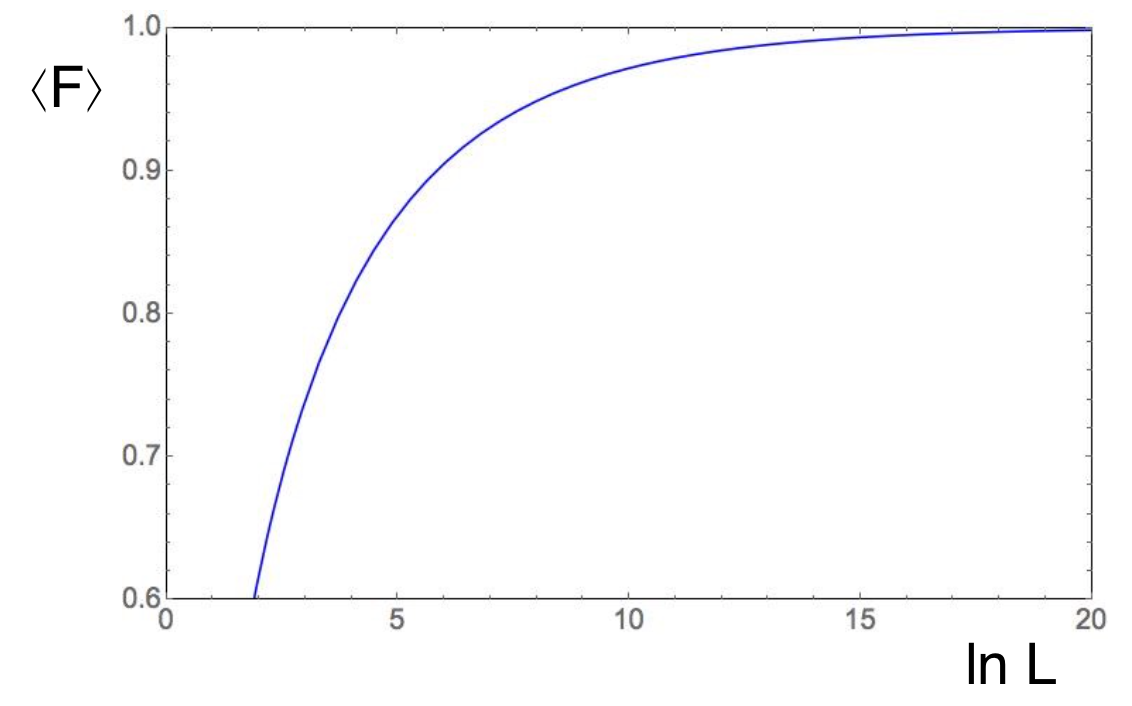}
\caption{The average fidelity $\langle F \rangle$ as function of 
the logarithm of  system size $ L$   at the 3D AMIT for $\eta = 2,$ $\lambda =1.$ }
\label{fig:averagef}
\end{figure}

We note that  the first moment of $I_A$
 yields the geometrical average of the fidelity
         $\exp (\langle \ln F \rangle) = \exp (-   \langle  I_A \rangle /2 )$
          giving a  typical value of $F$. 
        So far,  we have not yet obtained  the average fidelity $\langle F \rangle 
            = \langle \exp (-  I_A  /2 ) \rangle$, 
             since its calculation requires the knowledge of 
               all  moments of $I_A$ \cite{gefencomment}.
                Using the distribution function as obtained in the pair approximation 
                 above, Eq. (\ref{p}), we find at the AMIT
                 \begin{equation} \label{averagef}
             \langle F \rangle = \frac{1}{\sqrt{8 \pi \eta \ln L}}  \int_{-\infty}^{\infty} d x 
             e^{  -\langle I_A \rangle L^{-\eta} \exp (x) }
         e^{      - \frac{x^2}{8 \eta \ln L}}.
                 \end{equation}
                 An expansion in moments of $I_A$  gives a divergent series. 
                A  saddle point approximation to the integral 
                   in Eq. (\ref{averagef}), yields  $\langle F \rangle|_{L \rightarrow \infty}
                   = 1,$ as confirmed by numerical integration, 
                  Fig. \ref{fig:averagef}, where we plot  $\langle F \rangle$, Eq. (\ref{averagef}),
                     as function of $\ln L.$ 

We conclude that, typically, the Anderson orthogonality catastrophe 
 becomes  exponentially enhanced  at the AMIT.
  The typical fidelity  decays
 exponentially with system size $L$ as $F \sim \exp(-c L^{\eta})$, where $\eta $
   is the power of multifractal intensity correlations. 
    On  the metallic side of the transition we recover 
     Anderson's result that the typical overlap 
       decays with a power law $F \sim L^{-q (E_F)}$.
     The power increases as Fermi
         energy $E_F$ 
          approaches 
           mobility edge $E_M$ like $q (E_F) \sim  (\frac{E_F-E_M}{E_M})^{-\nu \eta},$
  where $\nu$ is the critical exponent of  correlation length $\xi_c$. 
     On the insulating side of the transition the typical value of  $F$
      approaches a constant for $L$ exceeding  localization 
       length $\xi$. While these results  for $F$ were obtained with the  
        mean values of Anderson integral $I_A$, we derive also its distribution 
          and find that $I_A$ is widely, log normally, distributed 
          with a width which diverges at the AMIT. 
           Surprisingly, we find that the average fidelity converges  at the AMIT  to
            $F=1$ as the system size $L$ is sent to infinity. This is
             a consequence of multifractality:  
              placing the additional short ranged impurity randomly in the sample, 
               the fidelity is typically exponentially small. 
               Averaging the 
                 fidelity,   there is a large weight on positions where the 
                 wave function intensity is reduced and where    the local density of states
              has   a local pseudogap.
At these positions the  additional impurity 
                   has no effect, so that $F=1$. As a consequence,
                    the average fidelity is  $  \langle F \rangle  \rightarrow 1$, while  typically
                    $ \exp (  \langle \ln  F \rangle ) \rightarrow 0$  at the AMIT 
                    in the infinite volume limit.          
        Building on  these  results we can next employ the same strategy to 
         study experimental 
         consequences like  singularities in  X-Ray absorption and emission
        \cite{nozieresdominicis} and the 
        zero bias anomaly  at  the MIT in doped semiconductors\cite{loehneysen}.  
We  also plan to  extend this approach  to explore the effect of 
  more extended impurities\cite{gefenlerner}, other 
  parametric perturbations \cite{venuti} and to the study of 
   nonequilibrium quantum dynamics of disordered systems which can
   be studied in synthetic many-body  systems with  controlled disorder
  in ensembles of ultracold atoms.


\acknowledgments

We gratefully acknowledge Yuval Gefen and Vadim 
Cheianov for stimulating  and usefull 
discussions and Stephan Haas for critical reading and useful comments. This research was supported by DFG grant KE 807/18-1. SK thanks the
Kavli Institute for Theoretical Physics, Beijing, for its  hospitality,
 where this work has been initiated during the workshop {\it Recent progress and perspectives in topological insulators}, organized by Victor Kagalovsky, Alexander L. Chudnovskiy, Sergey Kravchenko, Xin-Cheng Xie and Sen Zhou.



\end{document}